\documentclass[a4paper,11pt]{article}
\usepackage[toc,page]{appendix}
\usepackage{graphicx}
\usepackage{colortbl}
\usepackage{amsmath}
\usepackage{subfigure}
\usepackage{mathtools}
\usepackage[utf8]{inputenc}
\usepackage{float}
\usepackage{epstopdf}
\usepackage{amsfonts,amsmath,amssymb}
\usepackage{hyperref}

\usepackage{epsfig,multicol,bbm}

\newcommand\fverb{\setbox\pippobox=\hbox\bgroup\verb}
\newcommand\fverbit{\egroup\item[\fbox{\unhbox\pippobox}]}

\newbox\pippobox

\begin{document}
\title{\bf Gravitational Lensing by Polytropic Wormholes}
\author{S.N. Sajadi\thanks{Electronic address: naseh.sajadi@gmail.com}\,,\, N. Riazi\thanks{Electronic address: n\_riazi@sbu.ac.ir}
\\
\small Department of Physics, Shahid Beheshti University, G.C., Evin, Tehran 19839,  Iran}
\maketitle
\begin{abstract}
We obtain static multi-polytropic wormhole solutions in the framework of GR gravity. The resulting metric is asymptotically Minkowskian, and locally that of a wormhole. We also examine gravitational lensing by the wormhole, and calculate the deflection angle for weak and strong field limits. We investigate microlensing in the weak field limit and obtain corresponding light curves for both galactic and extragalactic situations. We discuss the multi-polytropic equation of state for the energy-momentum tensor which supports this geometry and finally, we check for the satisfaction of the weak energy condition.
\end{abstract}
\section{Introduction}
In classical general relativity, the Einstein field equations admit a simple and interesting class of static solutions describing tunnels in spacetime, connecting either two remote regions of our Universe or even different universes. These geometrical objects are called wormholes. The concept of a wormhole was first suggested by Ludwig Flamm by looking at the simplest solution of Einstein’s field equation (i.e. the Schwarzschild solution), in 1916, shortly after the publication of the Einstein field equations. By using the time symmetry of the solution, Flamm realized that Einstein equations allowed a second solution, now known as a white hole \cite{1}. In 1935, a similar construction was explored by Einstein and Rosen in a paper whose actual purpose was to explain fundamental particles such as electrons, in terms of spacetime tunnels threaded by electric lines of force \cite{2}, the so-called Einstein-Rosen bridge \cite{3}. The concept of “traversable wormhole” was suggested by Morris and Thorne with the imaginative idea of using a wormhole for interstellar travel or even time travel \cite{4}. In wormhole physics, one usually adopts the reverse problem of solving the Einstein field equation, by first constructing the spacetime metric, then deducing the stress-energy tensor components. It is readily found that traversable wormholes possess a stress-energy tensor that violates the null energy condition. In fact, they violate all the known pointwise energy conditions and averaged energy conditions, which are fundamental to the singularity theorems and theorems of classical black hole thermodynamics. In the opposite direction, if the exotic matter is removed from the wormhole throat, then the wormhole could decay into a black hole. It has been shown that the wormhole region that requires exotic matter can be made arbitrarily small by introducing a special shape function \cite{5}. It is remarkable that wormhole configurations may be constructed without needing any form of exotic matter sources in the framework of alternative theories of gravity, such as $ f(R) $ gravity,  Einstein-Gauss-Bonnet theory, Lovelock models, Brans-Dicke theory, among others \cite{6, 7, 8, 9, 10, 11, 12, 13}. In spite of several theoretical investigations, studies searching for observational evidence of the existence of wormholes are rare. Only a few attempts have been made to show the existence or nonexistence of wormholes in our universe \cite{14, 15, 16, 17}. A possible observational method that has been proposed to detect or reject the existence of wormholes is the application of gravitational lensing, since the light ray propagation is sensitive to local spacetime geometry.\\
From an observational point of view, lensing was the first observational verification of GR through investigation of starlight bending around the Sun during an eclipse in 1919 and continues to be a major source of astrophysical information. Gravitational lensing is an important tool to look for astrophysical objects such as black holes and dark matter halos. Early works focused on the lensing phenomenon in the weak field, but such a simplified analysis can not distinguish between different solutions that are asymptotically flat. For this reason, one must check the strong limit of gravitational lensing. The relation between the deflection angle and the properties of gravitational source close to a compact and massive source is known in integral form. The difficulty is that, light deflection diverges at the photon sphere (corresponding to unstable circular orbit close enough to the object such that it can loop any amount of times before reaching the observer) \cite{18}. Progress in this direction was initiated by Fritelli, Kling and Newman \cite{19}, and by Virbhadra and Ellis \cite{20}. When the light goes very close to a heavy compact body like a black hole, it will provide more information about the nature of the local gravitational field.

 Cramer et al. \cite{14} and Torres et al. \cite{15} have recently studied the microlensing effects of a negative mass wormhole on the light from a background point-like source. These authors have shown that the typical light curves expected from microlensing events produced by wormholes are very different from the ordinary positive mass microlensing of point sources. Also, these authors showed that the effective gravitational repulsion of light rays creates two asymmetric bursts. Recently, Anchordoqui et al. \cite{21} searched in the existing gamma-ray burst databases for signatures of wormhole microlensing. Although they detected some interesting candidates, no conclusive results were claimed \cite{ 21, 22}.\\

In the present work, we introduce a new exact wormhole metric which is supported by a multi-polytropic matter source. Polytropic equations of state are important in a variety of astrophysical applications. Traditionally, the polytropic equation of state has been used to describe a completely degenerate gas in white dwarfs and a completely convective star \cite{23}-\cite{ 31}. Moreover, different parts of an astrophysical objects may obey polytropic behavior, but with different polytropic indices. It is therefore reasonable to assume that an equation of state which is a mixture of different polytropes (i.e. a multi-polytropic EoS) can be a better fit to the real EoS. Considering a multi-polytropic EoS which is softer at low densities and stiffer at high densities may be appropriate for describing a hybrid astrophysical object. After all, a more general EoS provides a better fit, thereby increasing the freedom to describe naturally occurring astronomical objects \cite{32}. In Section $ 2 $, the general wormhole formalism is presented. In section $ 3 $ the formalism of gravitational lensing and microlensing is reviewed. In section $ 4 $, we solve the Einstein equations for multi-polytropic, spherical wormhole and calculate the strong lensing deflection angle numerically. In section $ 5 $, we investigate the corresponding energy-momentum tensor and discuss the energy conditions. The last section is devoted to conclusions and closing remarks.

\section{Wormhole Formalism}
 In the pioneering work of Morris and Thorne, they described a general class of
solutions of Einstein equations representing wormholes \cite{4}. The conditions that they imposed
in order to obtain the general form of the metric tensor included:\\
1. The metric must be spherically symmetric and static.\\
2. The solution must have a throat joining two asymptotically flat regions of spacetime.\\
3. The metric must satisfy Einstein equations in every point of spacetime.\\
4. The geometry will not have event horizons nor singularities.\\
Under these conditions, the general form for the metric is:
\begin{equation}\label{me1}
ds^{2}=-e^{2\Phi(r)}dt^{2}+\dfrac{dr^{2}}{1-\dfrac{b(r)}{r}}+r^{2}(d\theta^{2}+r^{2}\sin^{2} (\theta)d\phi^{2})
\end{equation}
where $ \Phi(r) $ and $ b(r) $ are arbitrary functions of the radial coordinate $ r $. $ \Phi(r) $ is called the redshift function, for it is related to the gravitational redshift, and $ b(r) $ is known as the shape function, because as can be shown by embedding diagrams, it determines the shape of the wormhole. The coordinate $ r $ is non-monotonic in that it decreases from $ +\infty $ to a minimum value $ r_{0} $, representing the location of the throat of the wormhole, where $ b(r_{0})=r_{0} $, and then it increases from $ r_{0} $ to $ +\infty $. The proper circumference of a circle of fixed $ r $ centered at the wormhole is given by $ 2 \pi r $, with the minimum value $ 2 \pi r_{0} $ at the throat. Although the metric coefficient $ g_{rr} $ becomes divergent at the throat, which is a coordinate singularity, the proper radial distance
\begin{equation}
l(r)=\pm\int _{r_{0}} ^{r}\dfrac{dr}{(1-\frac{b(r)}{r})^{\frac{1}{2}}}
\end{equation}
should be finite. Note that since $ 0 \leq 1-b(r)/r \leq 1 $, the proper distance is greater than or equal to the coordinate distance, i.e., $ |l(r)| \geq r-r_{0} $. The metric (\ref{me1}) may be written in terms of the proper radial distance as
\begin{equation}
ds^{2}=-e^ {2\Phi(l)}dt^{2}+dl^{2}+r^{2}(l)[d\theta^{2}+\sin^{2} (\theta)d\phi^{2}],
\end{equation}
The proper distance decreases from $  l = +\infty $, in the upper universe, to $ l = 0 $ at the throat, and then from zero to $ -\infty $ in the lower universe. For the wormhole to be traversable, it must have no horizons, which implies that $ g_{tt}=-\exp(2\Phi(r))\neq0 $, so that  $ \Phi(r) $ must be finite everywhere.
We can use embedding diagrams to represent a wormhole and extract some useful information for the choice of the shape function $ b(r) $. Due to the spherically symmetric nature of the problem, one may consider an equatorial slice, $ \theta=\frac{\pi}{2} $, without loss of generality. The respective line element, considering a fixed moment of time, $ t = const $, is given by
\begin{equation}\label{me2}
ds^{2}=\dfrac{dr^{2}}{1-\frac{b(r)}{r}}+r^{2}d\phi^{2}.
\end{equation}
To visualize this slice, one embeds this metric into three-dimensional Euclidean space, in which the metric can be
written in cylindrical coordinates, $ (r,\phi,z) $, as
\begin{equation}
ds^{2}=dz^{2}+dr^{2}+r^{2}d\phi^{2}
\end{equation}
Now, in the three-dimensional Euclidean space the embedded surface has equation $ z = z(r) $, and thus the metric of the surface can be written as,
\begin{equation}\label{me3}
ds^{2}=[1+(\frac{dz}{dr})^{2}]dr^{2}+r^{2}d\phi^{2}
\end{equation}
Comparing Eq. (\ref{me3}) with (\ref{me2}), we have the equation for the embedding surface, given by
\begin{equation}
\dfrac{dz}{dr}=\pm(\frac{r}{b(r)}-1)^{-\frac{1}{2}}.
\end{equation}
To be a wormhole solution, the geometry must have a minimum radius, $  r = b(r) = r_{0} $, denoted as the throat, at which the embedded surface is vertical, i.e., $ \frac{dz}{dr}\longrightarrow\infty $. Far from the throat this surface might be is asymptotically flat, $ \frac{dz}{dr}\rightarrow0 $ as $ r \rightarrow \infty $. One also needs to impose that the throat flares out. Mathematically,
this flaring-out condition means that the inverse of the embedding function $ r(z) $ must satisfy $ \frac{d^{2}r}{dz^{2}}>0 $ at or near the throat $ r_{0} $. Differentiating $ \frac{dr}{dz}=\pm(\frac{r}{b(r)}-1)^{\frac{1}{2}} $ with respect to $ z $, we have
\begin{equation}\label{eight}
\dfrac{d^{2}r}{dz^{2}}=\dfrac{b-b^{\prime}r}{2b^{2}}>0
\end{equation}
At the throat, we see that inequality (\ref{eight}) implies the condition $ b^{\prime}(r_{0}) < 1 $. This condition plays a fundamental role in the analysis of the violation of the energy conditions \cite{4, 33, 34}.
To gain some insight into the fluid supporting the wormhole, Morris and Thorne defined the dimensionless function
\begin{equation}
\zeta=\frac{\bar{P}-\rho}{\rho}=\dfrac{P_{r}+2P_{t}-3\rho}{3\rho}.
\end{equation}
The wormhole material is exotic, i.e., $\zeta > 0 $, and nonexotic if $\zeta < 0 $. $ \zeta  $ is thus called the exoticity parameter \cite{4, 6, 35}.
\section{Einstein Field Equations and the Equation of State}
We consider a static spherically symmetric space-time in the form of metric (\ref{me1}). The non-vanishing components of the Einstein tensor for our ansatz metric read
\begin{equation}
G^{t}_{t}=\dfrac{b^{\prime}}{r^{2}},
\end{equation}
\begin{equation}
G^{r}_{r}=-\dfrac{b}{r^{3}}+2(1-\dfrac{b}{r})\dfrac{\Phi^{\prime}}{r},
\end{equation}
and
\begin{equation}
G^{\theta}_{\theta}=G^{\phi}_{\phi}=(1-\dfrac{b}{r})[\Phi^{\prime\prime}+(\Phi^{\prime})^{2}-\dfrac{b^{\prime}r-b}{2r(r-b)}\Phi^{\prime}-\dfrac{b^{\prime}r-b}{2r^{2}(r-b)}+\dfrac{\Phi^{\prime}}{r}].
\end{equation}
We therefore need a diagonal energy-momentum tensor $ T^{\mu}_{\nu}=diag(-\rho,P_{r},P_{t},P_{t}) $, with
\begin{equation}
\rho(r)=-\dfrac{1}{8\pi G}\dfrac{b^{\prime}}{r^{2}},
\end{equation}
\begin{equation}\label{epr}
P_{r}=-\dfrac{1}{8\pi G}[\dfrac{b}{r^{3}}+2(1-\dfrac{b}{r})\dfrac{\Phi^{\prime}}{r}],
\end{equation}
and
\begin{equation}
P_{t}=\dfrac{1}{8\pi G}(1-\dfrac{b}{r})[\Phi^{\prime\prime}+(\Phi^{\prime})^{2}-\dfrac{b^{\prime}r-b}{2r(r-b)}\Phi^{\prime}-\dfrac{b^{\prime}r-b}{2r^{2}(r-b)}+\dfrac{\Phi^{\prime}}{r}].
\end{equation}
By taking derivative with respect to the radial coordinate $ r $ of Eq. (\ref{epr}), and after some algebra, we obtain the relativistic hydrostatic equation \cite{naseh}:
\begin{equation}
\dfrac{dP_{r}}{dr}=\frac{G(\rho+P_r)(M(r)+4\pi r^3
P_r)}{r^2(1-\frac{2GM(r)}{r})}-\frac{2(P_r-P_t)}{r}=(\rho+P_{r})\Phi^{\prime}-\dfrac{2}{r}(P_{r}-P_{t})
\end{equation}\\
where $M(r)$ is the gravitational mass inside a sphere of radius $r$
\begin{equation}\label{eq8}
M(r)=\int _{r_{0}} ^{r} 4\pi r^2 \rho dr.
\end{equation}
Here, we assume the multi-polytropic equation of state in the following form \cite{naseh}
\begin{equation}\label{mol}
P_{r}=A\rho +B\rho^{\frac{2n+2}{n+2}}\\\:\: ,\:\:\:\:\:
P_{t}=C\rho + D\rho^{\frac{2n+2}{n+2}}+E\rho^{\frac{3n+2}{n+2}}.
\end{equation}
Although this equation of state is of a very special type and lacks generality, it can still be a useful model for the following reasons. Even for nonrelativistic polytropic spheres which lead to the Lane-Emden equation, analytical solutions can be obtained only for few certain values of the polytropic index $ n $ ($ n=0,1,5 $). For other values of the index (e.g. $ n= 3 $ etc.), only numerical or series solutions for small $ r $ can be found. Now, the relativistic case (i.e. the TOV equation) which is much more complicated, it is impossible to obtain exact analytical solutions for an arbitrary equation of state \cite{23,27,28}. This is why we have only numerical models for real astrophysical objects. Having an exact analytical solution should be interesting by its own. The cost, of course, is deviating from having a more realistic model. \\
It is interesting that the equation of state (\ref{mol}) is exactly consistent with the following simple and power law expressions for the redshift and shape functions. The red shift function and the shape function are:
\begin{equation}\label{phib}
\Phi(r)=-\dfrac{1}{2}(\dfrac{r_{0}}{r})^{n}\:\:\:\: ,\:\:\:\: b(r)=r(\dfrac{r_{0}}{r})^{n}.
\end{equation}
So, the metric turns out to be:
\begin{equation}\label{me34}
ds^{2}=-e^{-(\frac{r_{0}}{r})^{n}}dt^{2}+\dfrac{dr^{2}}{1-(\frac{r_{0}}{r})^{n}}+r^{2}d\theta^{2}+r^{2}\sin ^{2}\theta d\phi^{2}.
\end{equation}
Using the above metric and the Einstein equations, we obtain the energy-momentum tensor components:
\begin{equation}\label{rh2}
\rho =-\frac{1}{8\pi G }T^{t}_{t}=-\frac{1}{8\pi G}[\dfrac{(\frac{r_{0}}{r})^{n}(n-1)}{r^{2}}],
\end{equation}
\begin{equation}\label{pr2}
 P_{r}=\frac{1}{8\pi G} T ^{r}_{r}=\frac{1}{8\pi G}[\dfrac{(n-1)(\frac{r_{0}}{r})^{n}}{r^{2}}-\dfrac{n}{r^{2}}(\dfrac{r_{0}}{r})^{2 n}],
 \end{equation}
 and
 \begin{equation}\label{pt2}
P_{t}=\frac{1}{8\pi G}T^{\theta}_{\theta}=\frac{1}{8\pi G}[\dfrac{-2 n(n-1)}{4 r^{2}}(\dfrac{r_{0}}{r})^{n}+\dfrac{n^{2}}{r^{2}}(\dfrac{r_{0}}{r})^{2 n}-\dfrac{n^{2}}{4 r^{2}}(\dfrac{r_{0}}{r})^{3 n}].
\end{equation}
Comparing these with equation (\ref{mol}), one can obtain the constants
\begin{equation}
A=-1,\hspace{0.2cm}B=-\dfrac{n}{r^{2}_{0}}\left( \dfrac{r^{2}_{0}}{1-n}\right)^{\frac{2n+2}{n+2}},\hspace{0.2cm}C=\dfrac{n}{2},\hspace{0.2cm}D=\dfrac{n^{2}}{r^{2}_{0}}\left( \dfrac{r^{2}_{0}}{1-n}\right)^{\frac{2n+2}{n+2}},\hspace{0.2cm}E=-\dfrac{n^{2}}{4r^{2}_{0}}\left( \dfrac{r^{2}_{0}}{1-n}\right)^{\frac{3n+2}{n+2}}. 
\end{equation}
It can be seen that the spacetime is asymptotically flat if $ n>0 $.  For $ n<0 $, we will have a closed space with less symmetries compared to the Einstein static universe.
For the wormhole to be traversable, $ \Phi(r)=-\dfrac{1}{2}(\dfrac{r_{0}}{r})^{n} $ must be finite everywhere which again demands $ n>0 $.
The equation for the embedding surface is given by
\begin{equation}
\dfrac{dz}{dr}=\dfrac{1}{\sqrt{(\frac{r}{r_{0}})^{n}-1}}.
\end{equation}
At the throat, $ r=r_{0} $, the embedding surface is vertical i.e: $ \dfrac{dz}{dr}\longrightarrow\infty $.
Far from the throat, the space is asymptotically flat, i.e  as $ r\longrightarrow\infty $ , $\dfrac{dz}{dr}\longrightarrow0 $.\\
In order to check the flare-out condition at the throat, we have
\begin{equation}
\dfrac{d^{2}r}{dz^{2}}=\dfrac{n}{2r}(\dfrac{r}{r_{0}})^{n},
\end{equation}
and since the shape function satisfies the condition $ b^{'} (r_{0})<1$, we should have (using eq. (\ref{rh2}))
\begin{equation}
b^{'}(r_{0})=(1-n)(\dfrac{r_{0}}{r})^{n}\vert_{r=r_{0}}=1-n<1
\end{equation}
which requires $ n>0 $ once again. The gravitational mass $M(r)$ inside a sphere of radius $r$ is
\begin{equation}\label{eq41}
M(r)=\int _{r_{0}} ^{r} 4\pi r^2 \rho dr=4\pi r_{0}[-1+(\dfrac{r_{0}}{r})^{n-1}]
\end{equation}
In order to have positive mass, we must have $ n<1 $, while, in the case of $ n=1 $, we have massless wormhole.
Therefore, in order to satisfy the above conditions (flare-out, asymptotic flatness and positive mass), we should have $0< n<1 $.
\section{Gravitational Lensing and Microlensing by Wormholes}
Gravitational lensing occurs when a gravitating mass distorts spacetime
and the geodesic of anything passing nearby it. The paths followed by electromagnetic radiation from
a star, galaxy, or other sources are bent as well. This can be seen directly from
equations of motion for photons. To show the effect of a gravitating mass on the path
of light, we consider a general spherically symmetric and static line element \cite{18,19,20}:
\begin{equation}\label{me10}
ds^{2}=-A(r)dt^{2}+B(r)dr^{2}+C(r)(d\theta^{2}+\sin^{2} (\theta)d\phi^{2}).
\end{equation}
Photons follow the equations of motion for null geodesics:
\begin{equation}
\dfrac{d^{2}x^{\mu}}{d\lambda^{2}}+\Gamma^{\mu}_{\alpha\beta}\dfrac{dx^{\alpha}}{d\lambda}\dfrac{dx^{\beta}}{d\lambda}=0
\end{equation}
where $ \lambda $ is the affine parameter and $ \frac{dx^{\mu}}{d\lambda} $ represents the four-velocity of the photons.
In a spherically symmetric spacetime, the lens, observer, and source can all be placed on a single plane with a constant value of $ \theta=\frac{\pi}{2} $, so  $ \dot{\theta} = 0 $. Without loss of generality the metric reduces to:
\begin{equation}
ds^{2}=-A(r)dt^{2}+B(r)dr^{2}+C(r)d\phi^{2}.
\end{equation}
Since the metric is independent of $ t $ and $ \phi $, there are two conserved quantities:
\begin{equation}\label{ar}
 A(r)\dot{t}=E \hspace{1cm},\hspace{1cm} C(r)\dot{\phi}=J.
\end{equation}
For the metric (\ref{me10}), the geodesic equation becomes:
\begin{equation}\label{gu}
-\dfrac{1}{u^{2}}+\dfrac{A(r)B(r)}{J^{2}}(\dfrac{dr}{d\lambda})^{2}+\dfrac{A(r)}{C(r)}=0\:\:\:,\:\:\:where\;\;
u=\dfrac{J}{E}=\sqrt{\dfrac{C(r_{m})}{A(r_{m})}}.
\end{equation}
We want to know how much $ \phi $ changes as the light ray moves from the left side to the right side. The light ray travels from large $ r $ to a minimum $ r_{m} $ and then out to large $ r  $ again. Since both the source and the observer are located very far from the lens and there is negligible deflection far away from the lens, we can treat both as located at $ r=\infty $. To calculate the change in $ \phi $ with $ r $ over the light ray’s trajectory, we perform the integral:
\begin{equation}
\alpha(r_{m})=2\int _{r_{m}} ^{\infty}\dfrac{d\phi}{dr}dr-\pi=2\int _{r_{m}} ^{\infty}\dfrac{\dot{\phi}}{\dot{r}}dr-\pi.
\end{equation}
where $ \alpha(r_{m})=\phi(+\infty)-\phi(-\infty) $ is the deflection of light due to a mass as a function of its point of closest
approach. This integral sums the deflection at each value of $ r $ from $ r_{m} $ to $ \infty  $. This can be interpreted as the deflection that occurs during light's trip from $ r_{m} $ to the observer. The integral is multiplied by two to take into account the deflection that takes place during its trip from the source to $ r_{m} $. By using Eqs. (\ref{ar}) and (\ref{gu}), we obtain:
\begin{equation}\label{li}
\alpha(r_{m})=\int _{r_{m}} ^{\infty}\dfrac{2\sqrt{B(r)}dr}{\sqrt{C(r)}\sqrt {\dfrac{C(r)A(r_{m})}{C(r_{m})A(r)}-1}}-\pi.
\end{equation}
Because this integration can not be performed exactly for most metrics, usually it is approximated in the weak and strong cases \cite{19, 36}.
By applying the above general result and using the limit $ \frac{M}{r_{m}}<<1 $ (weak field limit) to the Schwarzschild metric one obtains the following result
\begin{equation}\label{we}
\alpha(r_{m})=\dfrac{4M}{r_{m}}
\end{equation}
which is used for almost all astrophysical considerations of lensing.
In a strong field limit people commonly use Bozza$^{\prime}$s method to calculate lensing effect of objects \cite{19}. The method of Bozza is to expand the integral in Eq. (\ref{li}) around the photon sphere to obtain a logarithmic expression for $ \alpha(r_{m}) $ . According to that procedure the deflection angle is obtained as follows
\begin{equation}
\alpha(\theta)=-\bar{a} \ln{(\frac{\theta D_{OL}}{u_{m}}-1)}+\bar{b},
\end{equation}
where $ \bar{a} $ and $ \bar{b} $ are constants and $ u_{m} $, $ D_{OL} $ are the impact parameter of the light and the distances from the observer to the lens, respectively \cite{37}-\cite{ 41}.
The angle between the lens (wormhole) and the source $  \beta$ can then be written as
\begin{equation}\label{19}
\beta=\dfrac{\xi}{D_{L}}-\dfrac{D_{LS}}{D_{S}}\alpha(r),
\end{equation}
where $ D_{L} $, $ D_{S} $, $ D_{LS} $, and $ \xi $ are the distances from the observer to the lens, from the observer to the source, and from the lens to the source, and the impact parameter of the light ray, respectively.
If the source and lens are completely aligned along the line of sight, the image is expected to be circular (an Einstein ring). The Einstein radius $ R_{E} $, which is defined as the radius of the circular image on the lens plane, is obtained from Equation with $ \beta=0 $ as
\begin{equation}\label{radi2}
R_{E}=\sqrt{\dfrac{4GM}{c^{2}}\dfrac{D_{LS}D_{L}}{D_{S}}},
\end{equation}
When both the lens and the source are moving with respect to each other, then the image changes its position and brightness. This effect is called microlensing. Angle $ \beta $ changes with time according to
\begin{equation}
\beta(t)=\sqrt{\beta_{0}^{2}+(\frac{t-t_{0}}{t_{E}})^{2}},
\end{equation}
where $ \beta_{0} $ is the impact parameter of the source trajectory and $ t_{0} $ is the time of closest approach. $ t_{E} $ is the Einstein radius crossing time given by
\begin{equation}
t_{E}=\dfrac{R_{E}}{V_{T}},
\end{equation}
where $ V_{T} $ is the transverse velocity of the lens relative to the source and observer. Finally, the total amplification as a function of time $( t=t_{0}\pm t_{E}) $ is given by \cite{14,15}
\begin{equation}
A(t)=\dfrac{\beta(t)^{2}\pm2}{\beta(t)\sqrt{\beta(t)^{2}\pm4}},
\end{equation}
where the plus (minus) sign corresponds to the positive (negative) point mass case. When the source is extended, we must take into account the contributions coming from the different parts of it \cite{42}. If the lens is extended and source is point like, then we can take into account the mass of lens enclosed to region with radius of impact parameter \cite{43}, or characterize an extended lens by its virial mass: the mass contained in a sphere with mean density equal to the virial density \cite{44}. \\
\subsection{The case  $0<n<1$}\label{se4}
\begin{figure}[H]\hspace{0.4cm} 
\centering
\subfigure[]{\includegraphics[width=0.4\columnwidth]{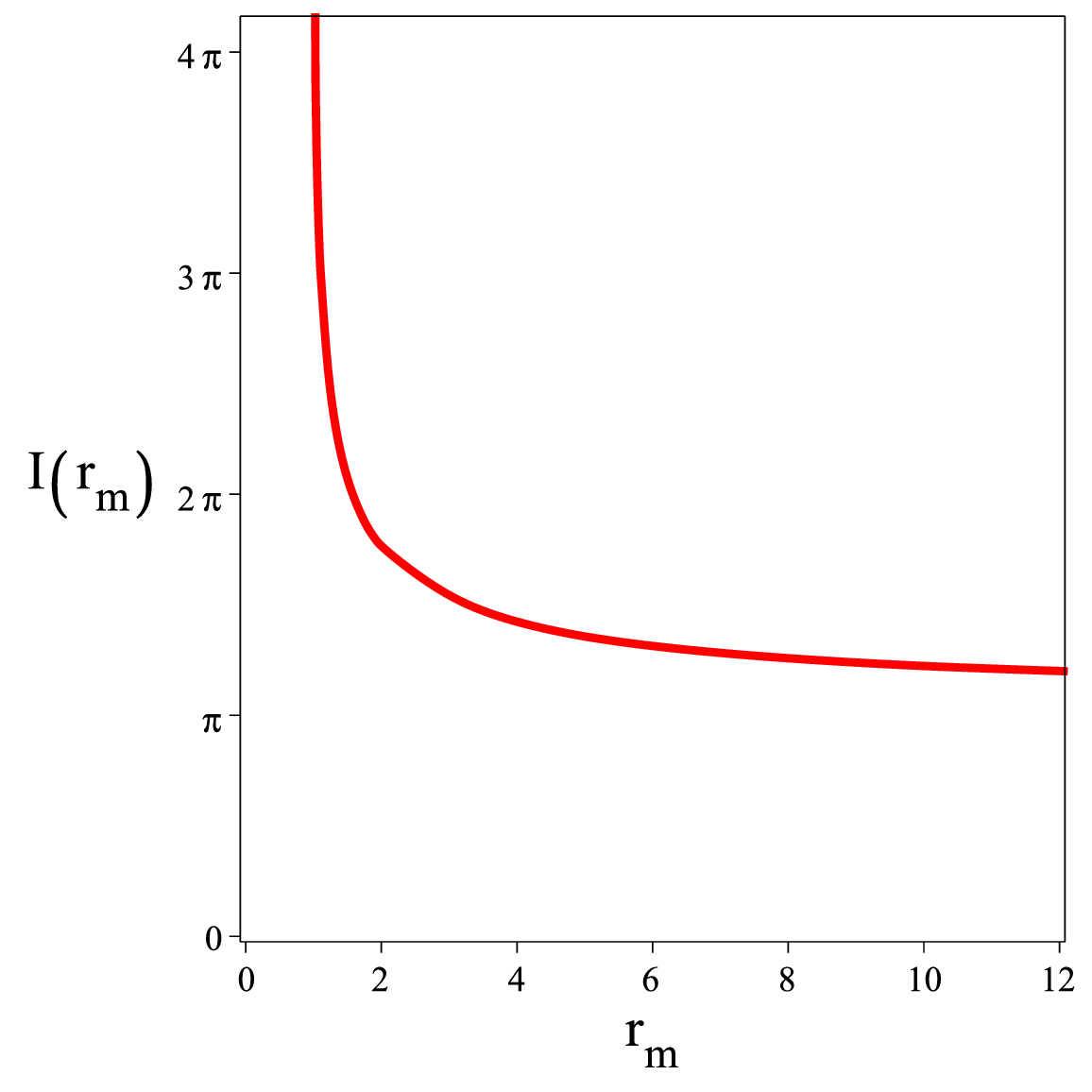}}
\subfigure[]{\includegraphics[width=0.4\columnwidth]{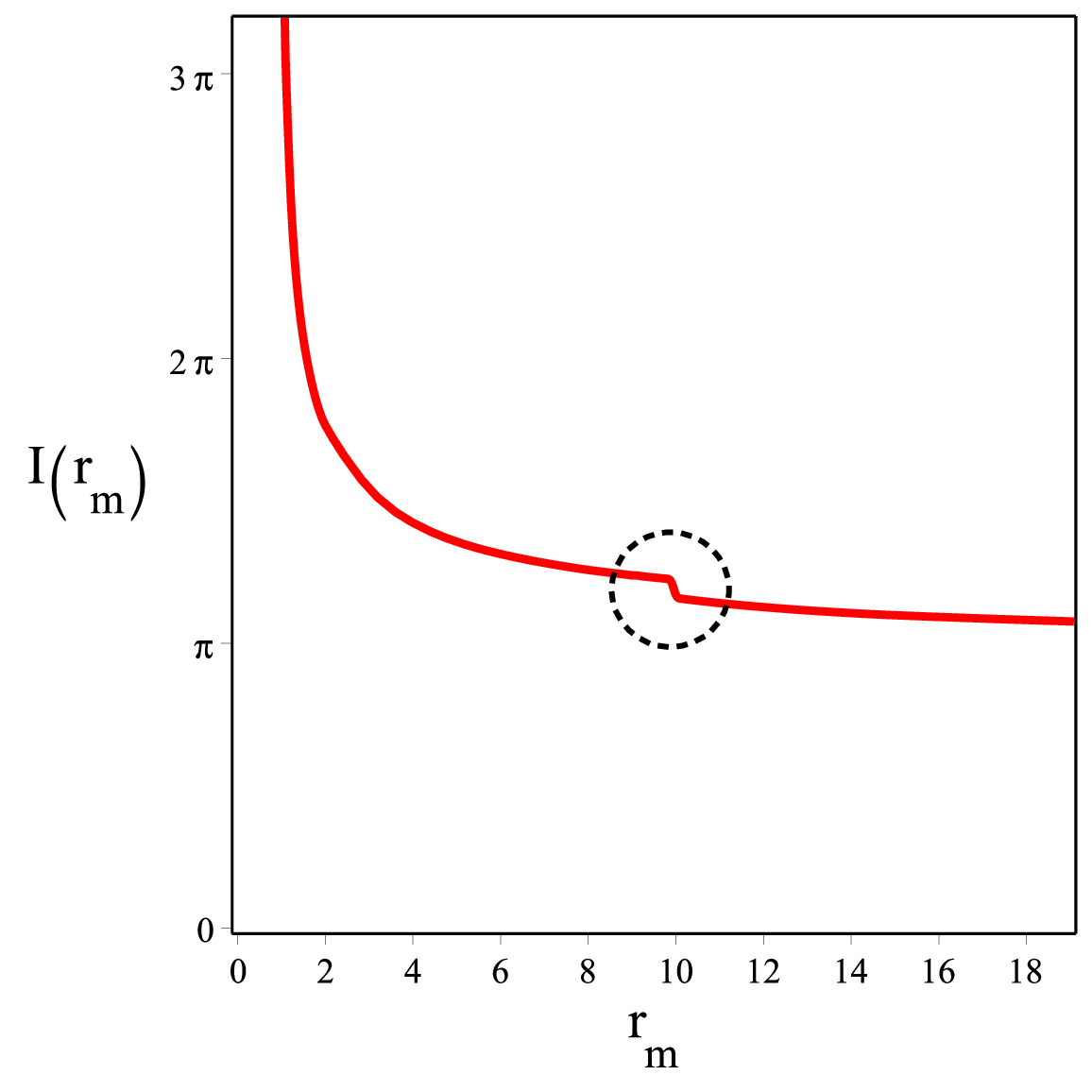}}
\vspace{0.2cm}
\caption{The behavior of $ I $ in terms of $ r_{m} $ for $ r_{0}=1$, $ n=\frac{1}{2} $, if the metric (\ref{me34}) is used all over the space (left). The behavior of $ I $ in terms of $ r_{m} $ for $ r_{0}=1,   m=1, n=\frac{1}{2}$ and linearly-interpolated metric for the thin shell (right).} \label{fig:2}
\end{figure}

\begin{figure}[H]\hspace{0.4cm} 
\centering
\subfigure[]{\includegraphics[width=0.4\columnwidth]{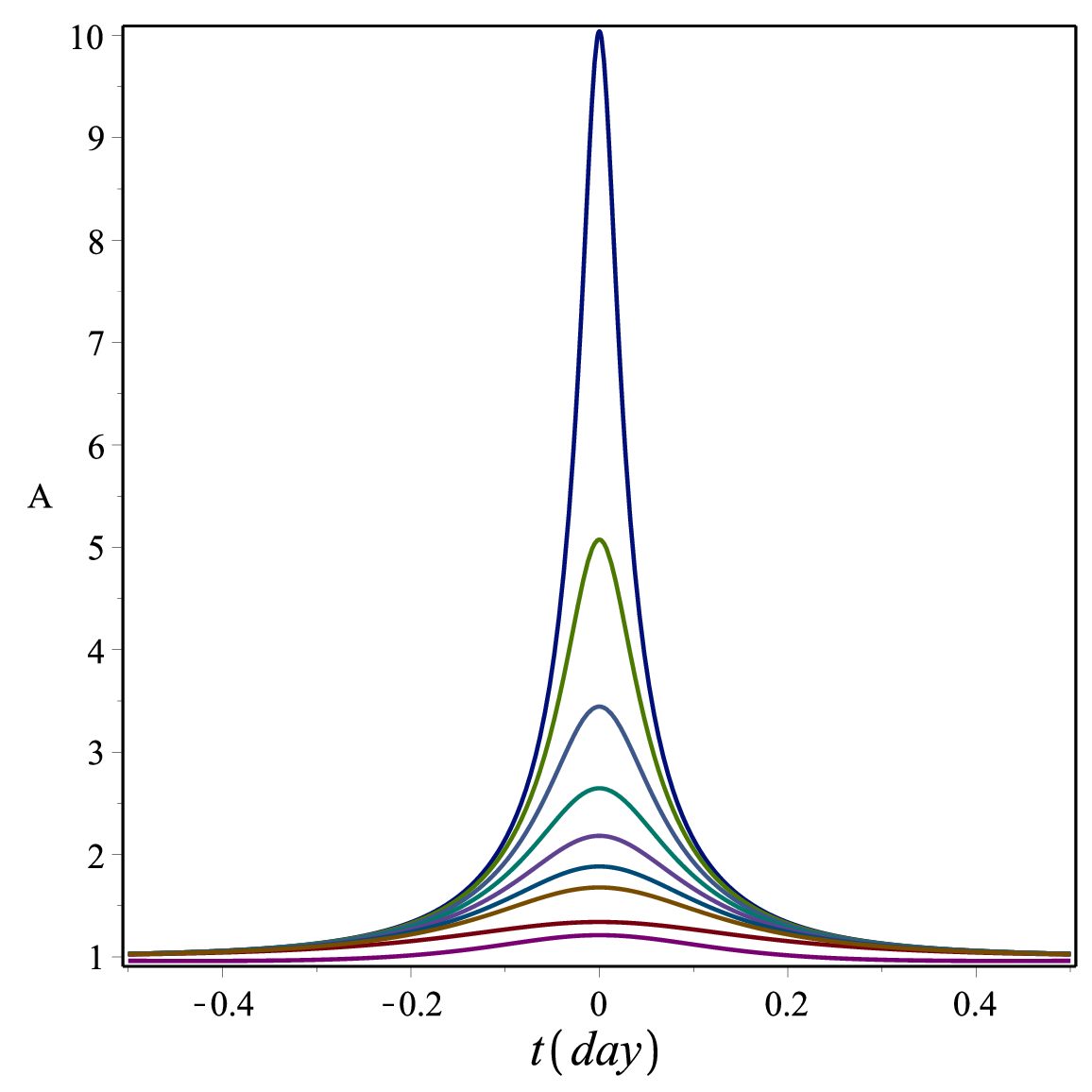}}
\subfigure[]{\includegraphics[width=0.4\columnwidth]{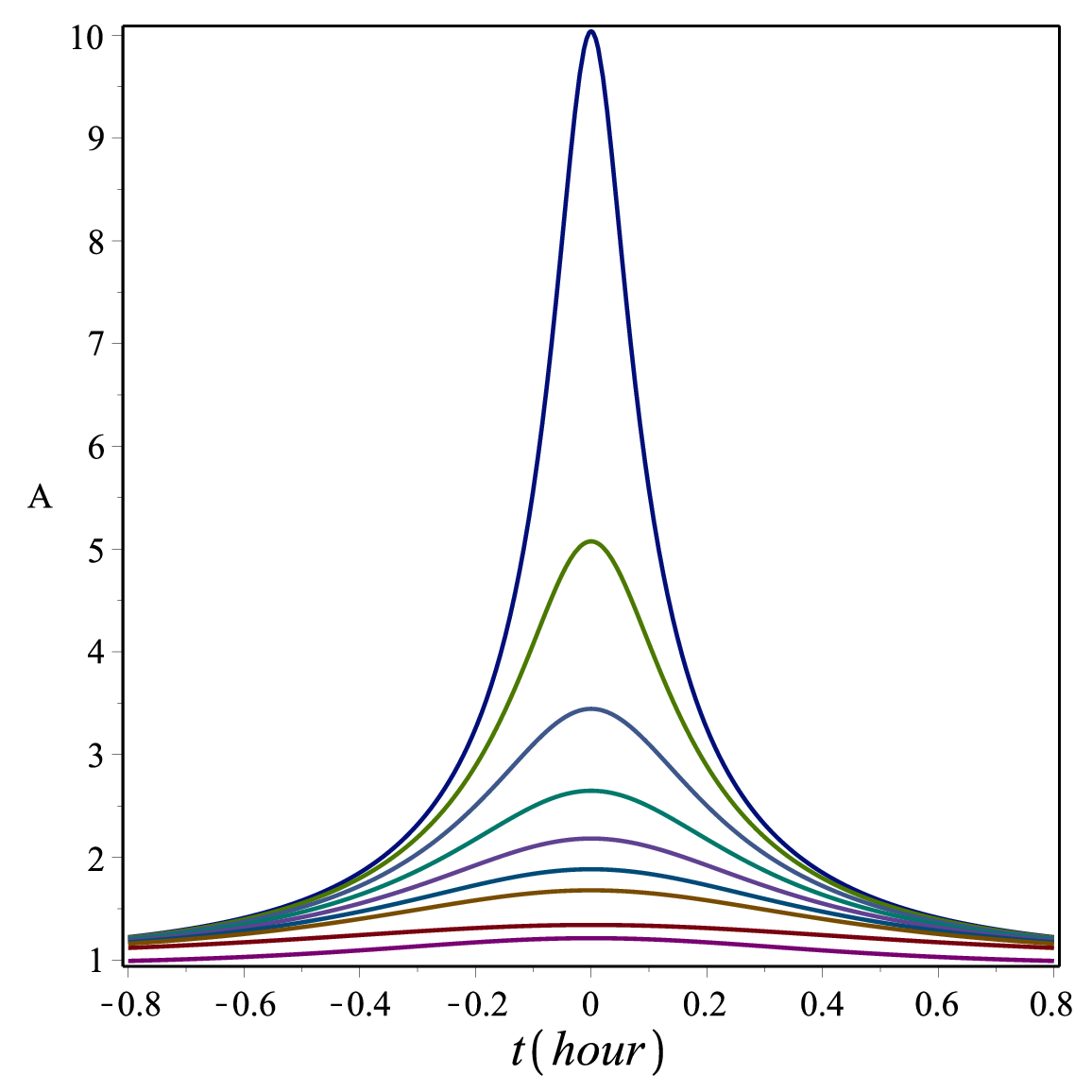}}
\vspace{0.2cm}
\caption{The behavior of amplification ($ A $) verses time ($ t $) for unbound wormhole at $ n=\frac{1}{2} $, $ r_{0}=1 $, $ \beta_{0}=0.1,0.2,0.3,0.4,0.5,0.6,0.7,0.8,0.9,1 $, $ V_{T}=220\frac{km}{s} $ (left) and $ V_{T}=5000\frac{km}{s} $ (right).} \label{fig4}
\end{figure}

\begin{figure}[H]
\begin{center}
\includegraphics[scale=0.5,width=0.5\textwidth]{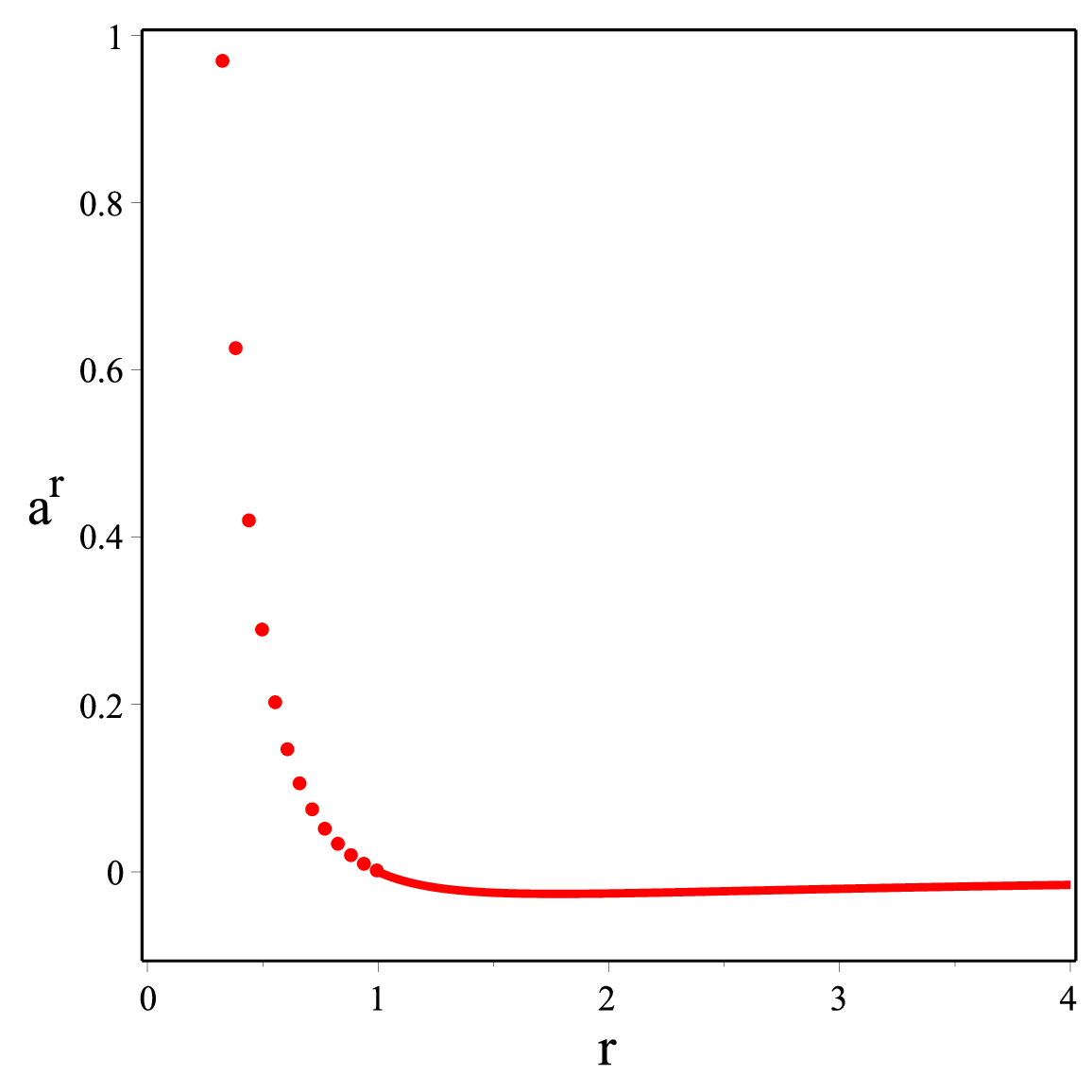}
\vspace{0.2cm}
\caption{\footnotesize{The behavior of $ a^{r}$ for $ n=\frac{1}{2} $ and $ r_{0}=1 $.}}\label{fig7}
\end{center}
\end{figure}

In this interval, the mass function is positive throughout the spacetime, but unbounded as $ r \rightarrow \infty $. In order to have a finite mass in this case, one has to put a thin shell layer and glue the wormhole spacetime to a Schwarzschild external spacetime \cite{46,47}.  
By using the wormhole metric and Eq. (\ref{li}) for deflection angle, one obtains the following quadrature for lensing, if the metric (\ref{me34}) is used all over the space:
\begin{equation}\label{lem}
\alpha(r_{m})=\int _{r_{m}} ^{\infty}\dfrac{2dr}{r((1-(\frac{r_{0}}{r})^{n})(\dfrac{r^{2}e^{(\frac{r_{0}}{r})^{n}}}{r_{m}^{2}e^{(\frac{r_{0}}{r_{m}})^{n}}}-1))^{\frac{1}{2}}}-\pi,
\end{equation}
where $ r_{m} $ is the coordinate of closest approach. If we rewrite the integral as
 \begin{equation}
 \alpha(r_{m})=I(r_{m})-\pi,
 \end{equation}
for a finite mass model, we use the relation described in section (\ref{se4}), and calculate $ I(r_{m}) $ by splitting the integral into two parts for $\left[  r_{0}, r_{1}\right] $ and $ \left[ r_{1}, \infty\right]  $. Each integral is evaluated by numerical methods. Results are plotted in Figure \ref{fig:2}.
Full gravitational lensing (both weak and strong) can be read from Fig. \ref{fig:2} and Eq. (\ref{lem}). According to this figure, the behavior is similar to the Schwarzschild metric which has positive deflection angle, i.e. wormhole acts like convergent lens. Note that lensing calculations which have led to Fig. \ref{fig:2}b are performed by assuming $r_{0} =1$ and $ r_{1} =10$. The divergence occurs in the throat and corresponds to the unstable photon sphere.
As we get far from the photon sphere, $ I(r_{m}) $ becomes smaller and eventually becomes $ \pi $, that means at long distances the deflection angle goes to zero which is consistent with weak lensing. 
Also, the effect of the junction layer is obvious in Fig. \ref{fig:2}. As seen, there is a jump in the light path.
This behavior can be explained from the viewpoint of Israel junction conditions or Fermat$ ^{\prime} $s principle. From the viewpoint of junction conditions, the Christoeffel symbols are discontinuous and this causes a jump in null geodesics at $ r_{1} $. From second fundamental form of the boundary, the extrinsic curvature for the two side of thin shell and with the intrinsic coordinates $ \xi^{i} $ on the shell is 
\begin{equation}\label{ext}
K_{ab}^{\pm}=-n^{\pm}_{\mu}\left( \dfrac{\partial^{2}x^{\mu}}{\partial\xi^{a} \partial\xi^{b}}+\Gamma^{\pm}{}^{\mu}_{\alpha \beta}\dfrac{\partial x^{\alpha}}{\partial \xi^{a}}\dfrac{\partial x^{\beta}}{\partial \xi^{b}}\right)_{\Sigma} 
\end{equation}
where $ n^{\pm}_{\mu} $ are given by
\begin{equation}\label{norm}
n^{\pm}_{\mu}=\pm\left( \dfrac{1}{\sqrt{g^{\alpha \beta}\dfrac{\partial \Phi}{\partial x^{\alpha}}\dfrac{\partial \Phi}{\partial x^{\beta}}}}\dfrac{\partial \Phi}{\partial x^{\mu}}\right)_{\Sigma}. 
\end{equation}
$ \Phi=0 $ is the parametric equation of $ \Sigma $ and $ - $ and $ + $ corresponds to two sides of the thin shell (for the present model $ \Phi=r-r_{1,2}=0 $). By using Lanczos equations, we have the surface energy $ \sigma $ and surface tangential pressures $ P $ on the thin shells by calculating 
\begin{equation}
S^{i}_{j}=-\dfrac{1}{8\pi}\left( \left[ K^{i}_{j}\right]-\delta^{i}_{j}\left[ K \right]\right),
\end{equation}
where $ [ K_{ij}]=K^{+}_{ij}- K^{-}_{ij} $ represent, the discontinuity
in the extrinsic curvature and for spherical symmetry case is $ [ K^{i}_{j}]=diag\left\lbrace[ K^{\tau}_{\tau}],[ K^{\theta}_{\theta}],[ K^{\phi}_{\phi}]\right\rbrace $ and $ S^{i}_{j}=diag(-\sigma, P_{\theta}, P_{\phi}) $ is the surface stress-energy tensor.
From Lanczos equations and equations (\ref{ext}) and (\ref{norm}) for a static metric one can obtain \cite{46}-\cite{47}
\begin{equation}
\sigma=-\dfrac{1}{4 \pi}\left[ K^{\theta}_{\theta}\right]= \dfrac{1}{8 \pi}\left[g^{\theta \theta}\partial_{r}g_{\theta \theta} \sqrt{g^{rr}}\right]_{\Sigma}=-\dfrac{1}{4\pi r_{1}}\left( \sqrt{1-\dfrac{2m}{r_{1}}}-\sqrt{1-\left( \dfrac{r_{0}}{r_{1}}\right)^{n}}\right)
\end{equation}
and
\begin{multline}
P_{\theta}=P_{\phi}=\dfrac{1}{8 \pi}\left( \left[ K^{\tau}_{\tau}\right]+\left[ K^{\theta}_{\theta}\right] \right)=\dfrac{1}{16 \pi}\left[\dfrac{1}{\sqrt{g_{rr}}}\left( \dfrac{\partial_{r}g_{\theta \theta}}{g_{\theta \theta}}+\dfrac{\partial_{r}g_{t t}}{g_{t t}}\right) \right]_{\Sigma}=\dfrac{1}{8 \pi}\left(\dfrac{\dfrac{m}{r^{2}_{1}}}{\sqrt{1-\dfrac{2m}{r_{1}}}}+\dfrac{1}{r_{1}}\sqrt{1-\dfrac{2m}{r_{1}}}\right)\\
-\dfrac{1}{8 \pi}\left( \dfrac{1}{r_{1}}\sqrt{1-\left(\dfrac{r_{0}}{r_{1}}\right)^{n}}-\dfrac{\sqrt{1-\left(\dfrac{r_{0}}{r_{1}} \right)^{n} }}{r_{1}}\dfrac{n}{2}\left( \dfrac{r_{0}}{r_{1}}\right)^{n}\right) .
\end{multline}
So, in the presence of a thin shell, one can match interior wormhole solution with the exterior Schwarzschild metric.\\
Here, we present an alternative approach which is based on the Fermat$ ^{\prime} $s principle. The gravitational refraction index depend on metric coefficients. To see this, we start from the action \cite{nst}   
\begin{equation}
A=\int d\lambda \sqrt{\dfrac{g_{ij}}{g_{tt}}\dot{X}^{i}\dot{X}^{j}}. 
\end{equation}
If we suppose $ g_{ij}=g(r)\delta_{ij} $ and $ g_{tt}=f(r) $, then the refraction index is as follows
\begin{equation}
n=\sqrt{\dfrac{g(r)}{f(r)}}.
\end{equation}
From Snell$^{'}$s law, the angles of incidence and refraction of light are related to each other according  to  
\begin{equation}\label{snel}
\dfrac{sin(\theta_{1})}{sin(\theta_{2})}=\dfrac{n_{2}}{n_{1}}.
\end{equation} 
For the present metric at boundary, $ n_{1} $ and $ n_{2} $ are 
\begin{equation}
n_{1}=\dfrac{1}{1-\dfrac{2m}{r_{1}}}\hspace{0.5cm},\hspace{0.5cm} n_{2}=\dfrac{1}{\sqrt{\left(1-\left(\dfrac{r_{0}}{r_{1}} \right)^{n}\right)e^{-\left( \dfrac{r_{0}}{r_{1}}\right)^{n} }} }
\end{equation}
for outer and inner sides, respectively. This leads to a breaking in the light trajectory across $ r_{1} $. For example, if $ r_{0}=1, \theta_{1}=\dfrac{\pi}{6}, r_{1}=10, m=1, n=\frac{1}{2} $ then $ \theta_{2}=\dfrac{\pi}{10} $.\\   
In the following, we consider microlensing of the wormhole. We will assume that $ \beta>>\alpha $ so that we are considering the weak-lensing regime. The energy density of wormhole is
\begin{equation}
\rho =\frac{1}{8\pi G}[ \dfrac{1-n}{r_{0}^{2}}(\dfrac{r_{0}}{r})^{n+2}].
\end{equation}
Since the wormhole is extended, the two-dimensional enclosed mass for an infinite spherical density is obtained as \cite{18}
\begin{equation}
M_{2D}=\dfrac{1}{4G}\pi^{\frac{1}{2}}[(1-n)r_{0}]\dfrac{\Gamma(\frac{n+1}{2})}{(1-n)\Gamma(\frac{n+2}{2})}(\dfrac{\xi}{r_{0}})^{1-n}
\end{equation}
for $ n=\frac{1}{2} $ and $ r_{0}=1 $ we have
\begin{equation}
M_{2D}=\dfrac{1}{G}\pi^{\frac{1}{2}}\dfrac{\Gamma(\frac{3}{4})}{\Gamma(\frac{1}{4})}\xi^{\frac{1}{2}}
\end{equation}
This is the mass enclosed by the cylinder interior to $ \xi $ and is obtained by integrating the projected surface mass density $ \Sigma $ over the area of the circle with radius $ \xi $. In a weak field, by using  Eq. (\ref{we}) for deflection angle and Eq. (\ref{19}), Einstein radius will be obtained
\begin{equation}\label{radi1}
R_{E}=(\dfrac{D_{L}D_{LS}\theta_{\alpha}r_{0}^{\frac{1}{2}}}{D_{S}})^{\frac{2}{3}},
\end{equation}
where $ \theta_{\alpha} $ is
\begin{equation}
\theta_{\alpha}=0.88 \mu as[\dfrac{\Gamma(\frac{3}{4})}{\Gamma(\frac{1}{4})}](1-\dfrac{D_{L}}{D_{S}})(\dfrac{pc}{16\pi G M_{\odot}}),
\end{equation}
in which we have assumed that the velocity of the wormhole is approximately equal to the rotation velocity of stars ($ V_{T}=220 \frac{km}{s}, D_{S}=8 kpc $ and $ D_{L}=4 kpc $) if it is bound to the Galaxy. If the wormhole is not bound to our Galaxy, the transverse velocity would be much higher. We assume ($ V_{T}=5000\frac{km}{s}, D_{S}=50 kpc $ and $D_{L}=25 kpc $)  for the unbound wormhole.
In Figure. \ref{fig4} we show the magnification curves and display two cases for two different velocity and different impact parameters. As one can see, the amplitude of light curves are similar to the light curves of the Schwarzschild point mass lenses, but the width of light curves are different. From the comparison of eq. (\ref{radi2}) and (\ref{radi1}) one can see that the width of wormhole light curves is larger compared to point mass lenses when impact parameter becomes small. Also as expected as the value of the impact parameter increases, the magnification becomes weaker.
Returning to the radial equation of motion (see Figure \ref{fig7}), for $ r>r_{0} $, radial acceleration is negative which means that the gravitational field of the wormhole is attractive, i.e, the free fall observer is pulled into the wormhole. Just at the throat $ r=r_{0} $, the radial acceleration and therefore, surface gravity is zero.
\subsection{Stability of the Thin Shell}
\begin{figure}[H]
\begin{center}
\includegraphics[scale=0.4,width=0.4\textwidth]{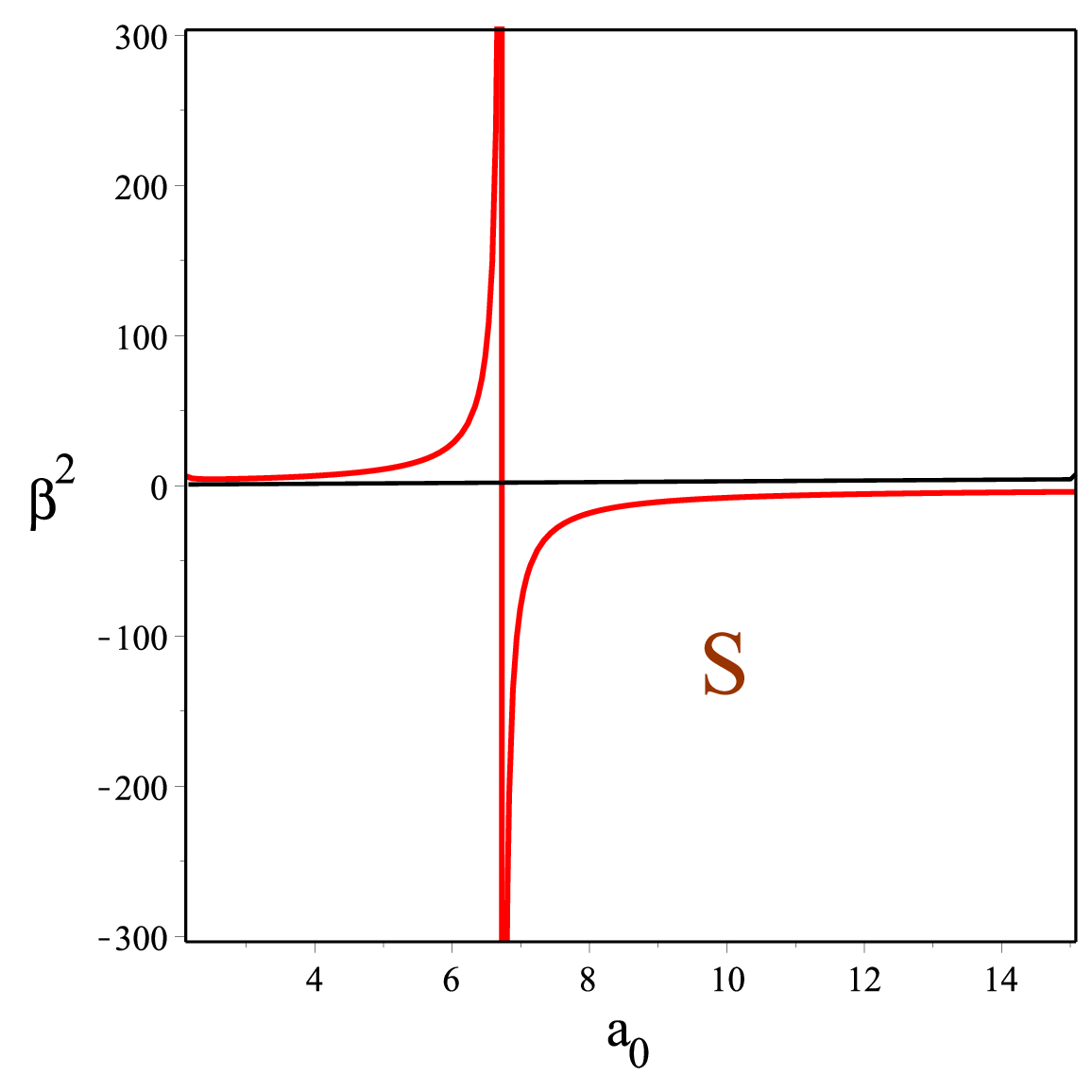}
\vspace{0.2cm}
\caption{\footnotesize{The behavior of $ \beta^{2} $ in terms of $ a_{0} $, at $ r_{0}=1,r_{1}=10,m=1,n=\frac{1}{2} $ .}}\label{figv}
\end{center}
\end{figure}

Here, we study the stability of the thin shell by linear perturbation of metric around a static solution. In order to study stability, we allow the radius of thin shell to become time dependent, $ r_{1}(\tau) $. One can show that the components of the extrinsic curvature in $ r_{1}(\tau) $ are given by \cite{46,47}
\begin{equation}
K_{\tau}^{\tau}{}^{+}=\dfrac{\dfrac{m}{r^{2}_{1}}+\ddot{r}_{1}}{\sqrt{1-\dfrac{2m}{r_{1}}+\dot{r}_{1}^{2}}},
\end{equation}

\begin{equation}
K_{\tau}^{\tau}{}^{-}=\dfrac{\sqrt{A+\dot{r}_{1}^{2}}}{A}\ddot{r}_{1}-\dfrac{n}{2r_{1}}\left( \dfrac{r_{0}}{r_{1}}\right)^{n}\dfrac{\sqrt{A+\dot{r}_{1}^{2}}}{A}\left( \dfrac{\dot{r}_{1}^{2}\left( \dfrac{r_{0}}{r_{1}}\right)^{n}-A^{2}}{A}\right)-\dot{r}_{1}^{2}e^{\dfrac{1}{2}\left( \dfrac{r_{0}}{r_{1}}\right)^{n}}\sqrt{A}\left(\dfrac{1+A+\dot{r}_{1}^{2}}{A+\dot{r}_{1}^{2}} \right)    ,
\end{equation}

\begin{equation}
K_{\theta}^{\theta}{}^{+}=\dfrac{1}{r_{1}}\sqrt{1-\dfrac{2m}{r_{1}}+\dot{r}_{1}^{2}},
\end{equation}

\begin{equation}
K_{\theta}^{\theta}{}^{-}=\dfrac{1}{r_{1}}\sqrt{1-\left( \dfrac{r_{0}}{r_{1}}\right)^{n}+\dot{r}_{1}^{2} },
\end{equation}
where $ A=1-\left( \dfrac{r_{0}}{r_{1}}\right)^{n}$. 
From Lanczos equation, one can obtain
\begin{equation}\label{sigm}
\sigma=-\dfrac{1}{4\pi r_{1}}\left( \sqrt{1-\dfrac{2m}{r_{1}}+\dot{r}^{2}_{1}}-\sqrt{1-\left( \dfrac{r_{0}}{r_{1}}\right)^{n} +\dot{r}^{2}_{1}}\right), 
\end{equation}
\begin{equation}
P=\dfrac{1}{8 \pi}\left(\dfrac{\dfrac{m}{r^{2}_{1}}+\ddot{r}_{1}}{\sqrt{1-\dfrac{2m}{r_{1}}+\dot{r}_{1}^{2}}}+\dfrac{1}{r_{1}}\sqrt{1-\dfrac{2m}{r_{1}}+\dot{r}_{1}^{2}}-\dfrac{1}{r_{1}}\sqrt{1-\left( \dfrac{r_{0}}{r_{1}}\right)^{n}+\dot{r}_{1}^{2} }-K^{\tau}{}^{-}_{\tau}\right). 
\end{equation}
Using equation (\ref{sigm}), one can obtain the equation of motion of the thin shell as
 \begin{equation}
 \dfrac{1}{2}\dot{r}_{1}^{2}+V(r_{1})=0,
 \end{equation}
 where potential is given by
 \begin{equation}
V(r_{1})=\dfrac{1}{2}\left(2-\dfrac{2m}{r_{1}}-\left( \dfrac{r_{0}}{r_{1}}\right)^{n}\right)+\dfrac{\left(\left( \dfrac{r_{0}}{r_{1}}\right)^{n}-\dfrac{2m}{r_{1}} \right)^{2} }{4\left( 4\pi r_{1}\sigma(r_{1})\right)^{2} } -\dfrac{\left(4\pi r_{1} \sigma(r_{1})\right)^{2}  }{4}.
\end{equation}
  In order to investigate the stability of the thin shell, one can expand $ V(r_{1}) $ around the static point $ r=a_{0} (\dot{a}_{0}=\ddot{a}_{0}=0) $ to obtain
 \begin{equation}
 V=V(a_{0})+V^{'}(a_{0})\left(r_{1}-a_{0} \right)+\dfrac{1}{2}V^{''}(a_{0})\left( r_{1}-a_{0}\right)^{2}+O\left(r_{1}-a_{0} \right)^{3}.   
 \end{equation}
 The first non zero term of the potential is $ V^{''}(a_{0}) $. So, the condition to stability is $ V^{''}(a_{0})>0 $ i.e. $ V(r_{1}) $ must have a minimum at $ a_{0} $.
 By using the conservation equation $ S^{i}_{j}{}_{;}{}_{i}+S^{i}_{\tau}{}_{;}{}_{i}=0 $ one can obtain
\begin{equation}\label{coi}
\sigma^{'}=-\dfrac{2}{r_{1}}\left(\sigma+P \right)+E 
\end{equation} 
 where 
\begin{multline}
E=\dfrac{S^{i}_{j}{}_{;}{}_{i}}{\dot{r}_{1}}=\dfrac{1}{\dot{r}_{1}}\left( T_{\mu \nu}U^{\mu}n^{\nu}\right)_{-}^{+}=\left(-T_{tt}\sqrt{\dfrac{1+g_{rr}\dot{r}_{1}^{2}}{g_{tt}}}+T_{rr}\sqrt{g_{rr}+g_{rr}^{2}\dot{r}_{1}^{2}} \right)_{-}^{+}\\
=\dfrac{\left( \dfrac{r_{0}}{r_{1}}\right)^{n}}{8\pi r_{1}^{2}}\left((1-n)e^{-\dfrac{1}{2}\left( \dfrac{r_{0}}{r_{1}}\right)^{n}}+\dfrac{\left( \dfrac{r_{0}}{r_{1}}\right)^{n}n-n+1}{\left(1-\left( \dfrac{r_{0}}{r_{1}}\right)^{n} \right)^{\dfrac{3}{2}} } \right)\hspace{5.5cm}.           
\end{multline}
Differentiating equation (\ref{coi}) with respect to $ r_{1} $ and using $ \beta^{2}=\dfrac{P^{'}}{\sigma^{'}} $, the following equation can be obtained
\begin{equation}
\beta^{2}=-1+\dfrac{r_{1}}{2\sigma^{'}}\left(E^{'}+\dfrac{2}{r_{1}^{2}}\left(\sigma+P \right)-\sigma^{''}\right). 
\end{equation}
As can be seen from Fig. \ref{figv}, the stable area is specified by $ S $ which corresponds to $ V^{''}>0 $. 

 \section{Energy Conditions and Exoticity Parameter}
\begin{figure}[H]
\begin{center}
\includegraphics[scale=0.4,width=0.4\textwidth]{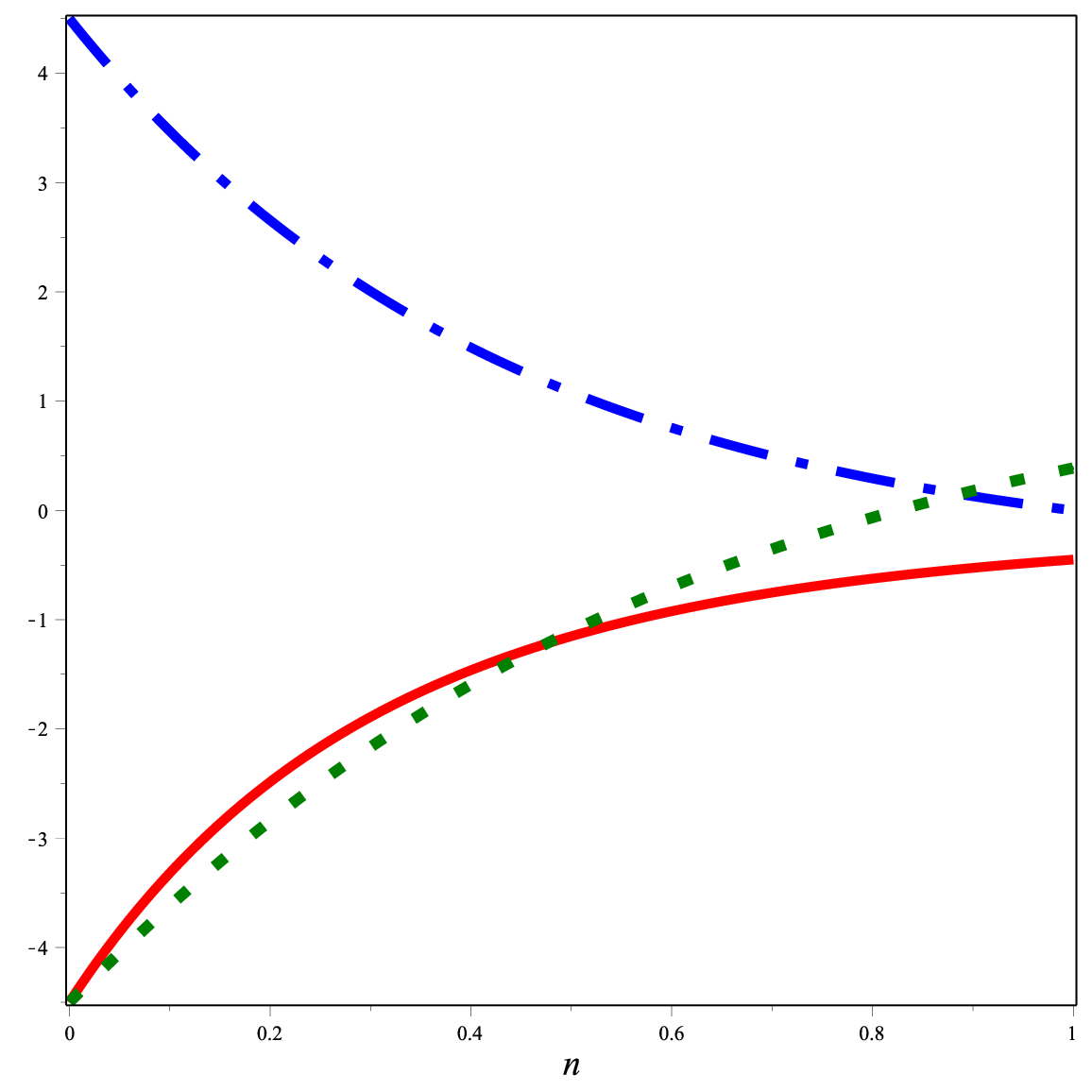}
\vspace{0.2cm}
\caption{\footnotesize{The behavior of average weak energy condition in terms of $ n $, $ r_{0}<r<10 r_{0} $.}}\label{fig8}
\end{center}
\end{figure}

As we mentioned in the introduction, the main objection against the plausibility of a wormhole
solution is that the energy-momentum tensor which supports this geometry violates the weak
energy condition (WEC). Since in some cases energy conditions are violated only locally (near to the throat), averaging of the energy conditions over timelike or null
geodesics would be a better global indicator. The averaged energy conditions are somewhat weaker than the pointwise energy conditions, as they permit localized violations of the energy conditions, as long as average energy conditions hold when integrated
along timelike or null geodesics \cite{4, 33}. As can be seen from Figure \ref{fig8}, the minimum of average energy density (\textcolor{blue}{dash dot  blue} color) is mainly positive, while the minimum of average $\rho+P_{r}$ (\textcolor{red}{ red} color) and  $ \rho+P_{t} $ (\textcolor{green}{dot green} color) are mainly below zero. So the minimum violation of weak energy condition are around $ n=0 $.

\begin{figure}[H]
\begin{center}
\includegraphics[scale=0.4,width=0.4\textwidth]{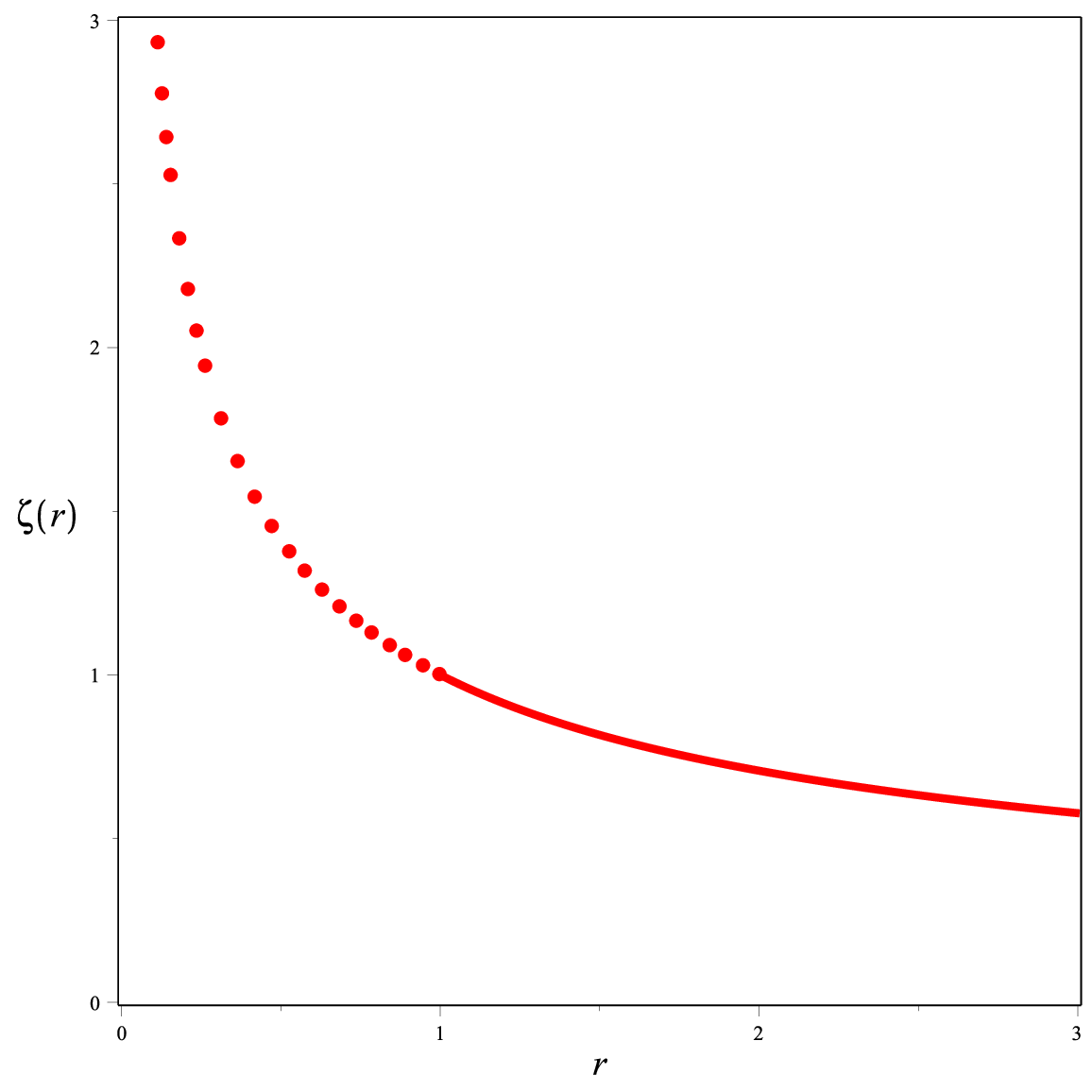}
\vspace{0.2cm}
\caption{\footnotesize{The behavior of $ \zeta(r)$ for $ n=\frac{1}{2} $ and $ r_{0}=1 $.}}\label{fig9}
\end{center}
\end{figure}
\begin{figure}[H]
\begin{center}
\includegraphics[scale=0.4,width=0.4\textwidth]{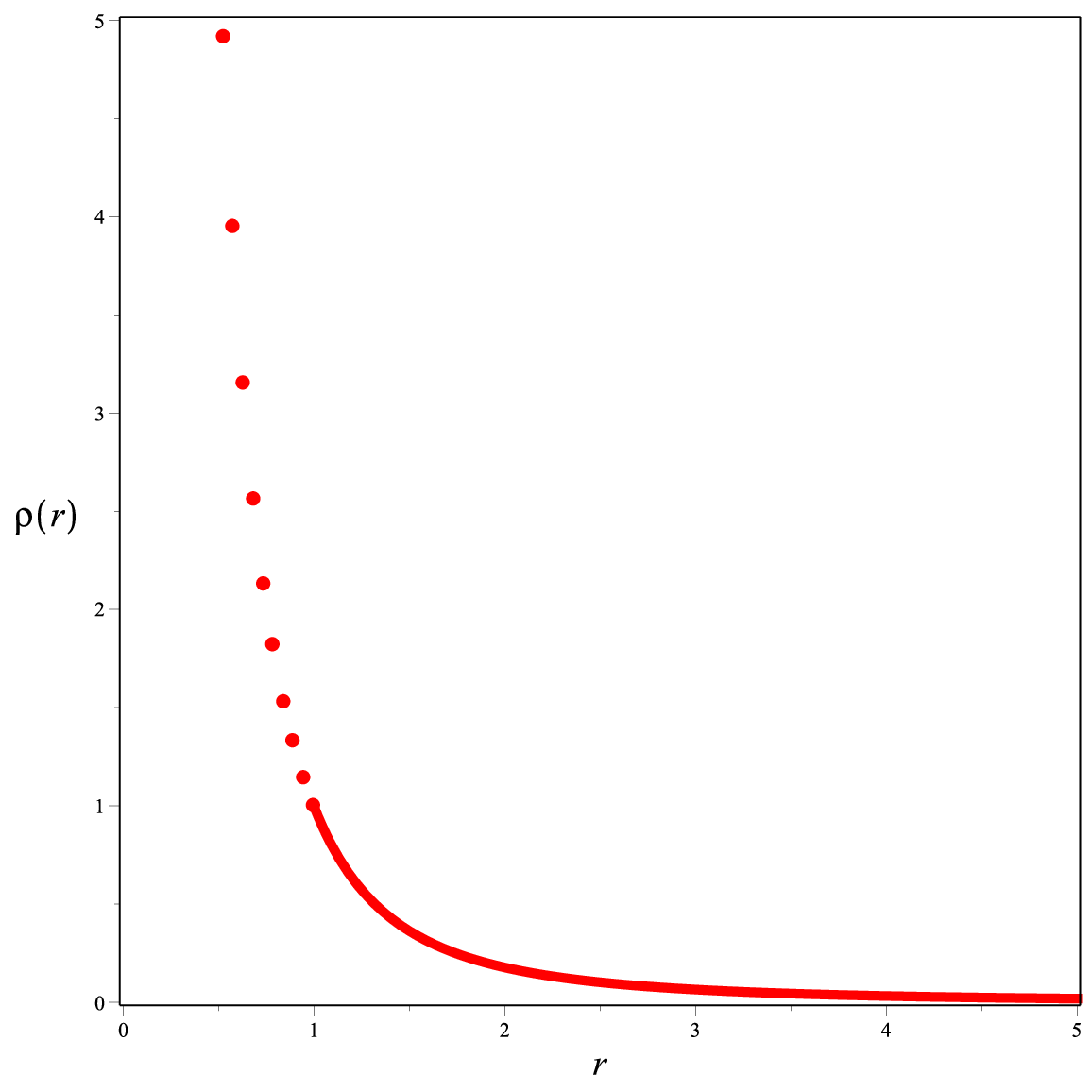}
\vspace{0.2cm}
\caption{\footnotesize{The behavior of $ \rho(r) $ for $ r_{0}=1, n=\frac{1}{2} $.}}\label{fig10}
\end{center}
\end{figure}
\begin{figure}[H]\hspace{0.4cm} 
\centering
\subfigure[]{\includegraphics[width=0.4\columnwidth]{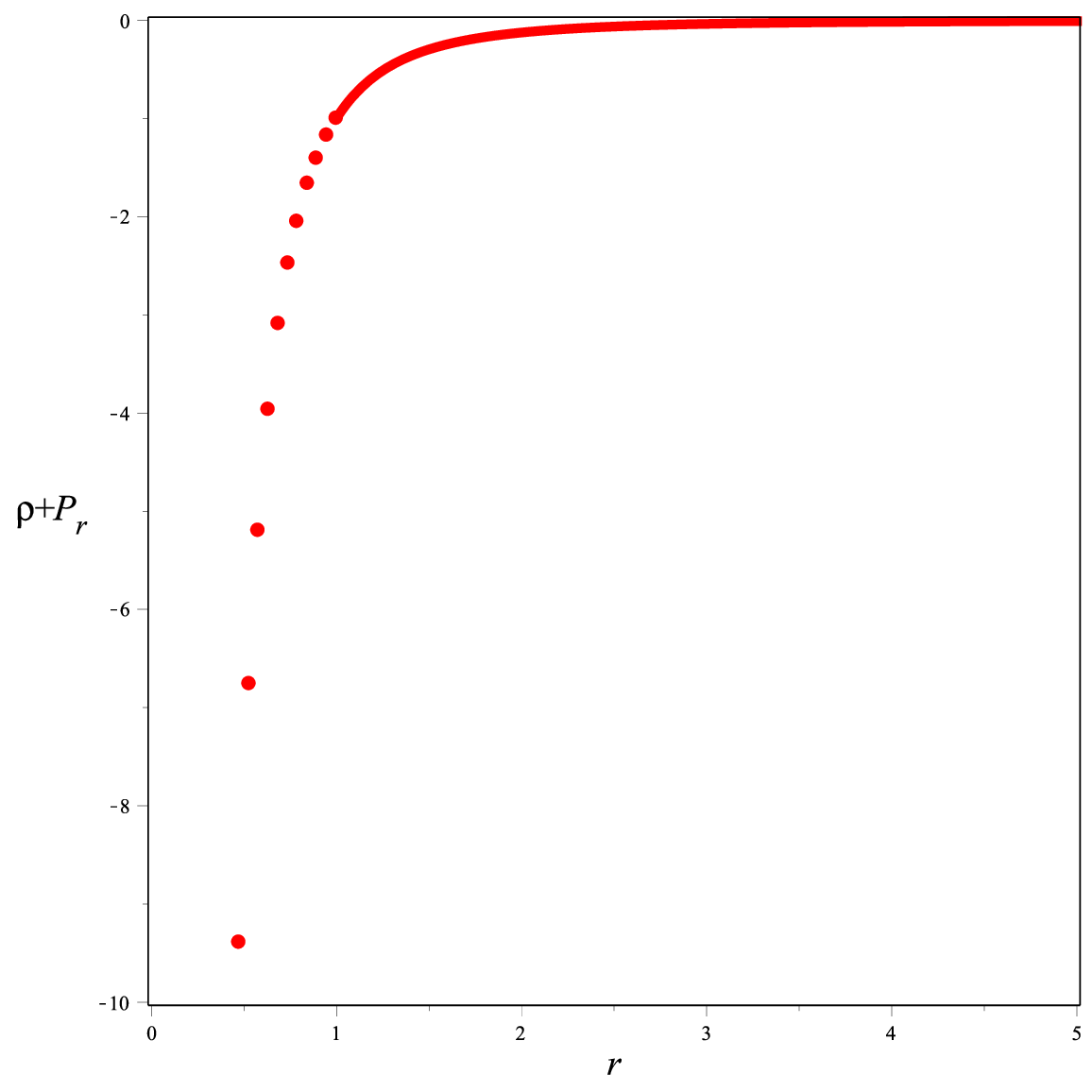}}
\subfigure[]{\includegraphics[width=0.4\columnwidth]{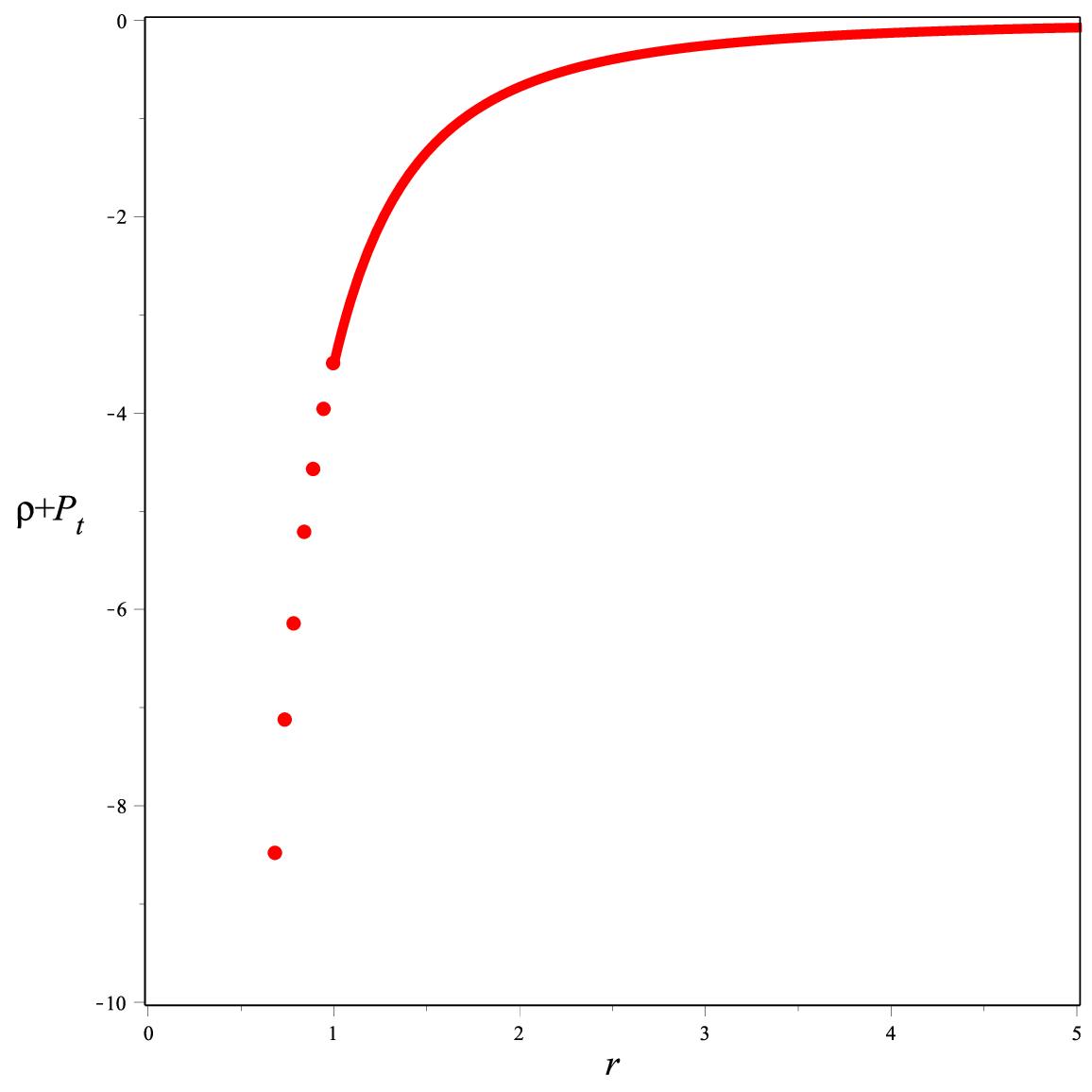}}
\caption{The behavior of $ \rho(r) +P_{r}(r)$ (left) and $ \rho(r)+P_{t}(r) $ (right) for $ r_{0}=1, n=\frac{1}{2}$.} \label{fig11}
\end{figure}

According to Figure \ref{fig9}, exoticity parameter is positive over the entire wormhole space.
 The weak energy condition requires
\begin{equation}
T_{\mu\nu}u^{\mu}u^{\nu}\geq0
\end{equation}
for every timelike $ u^{\mu} $ which leads to
\begin{equation}
\rho\geq0\:\:\:,\:\:\:\rho+P_{r}\geq0\:\:\:,\:\:\:\rho+P_{t}\geq0
\end{equation}
using (\ref{rh2}), (\ref{pr2}) and (\ref{pt2}) we have
\begin{equation}\label{rho}
\rho\geq0\:\:\:\rightarrow\:\: \dfrac{1-n}{r_{0}^{2}}(\dfrac{r_{0}}{r})^{n+2}\geq0
\end{equation}

\begin{equation}\label{pr}
\rho+P_{r}\geq0\:\:\:\rightarrow\:\:\: -\dfrac{n}{r_{0}^{2}}(\dfrac{r_{0}}{r})^{2 n+2}\geq0
\end{equation}
\begin{equation}\label{pt}
\rho+P_{t}\geq0\:\:\:\ \rightarrow\:\: \dfrac{(n+2)(n-1)}{2r_{0}^{2}}(\dfrac{r_{0}}{r})^{n+2}+\dfrac{n^{2}}{r_{0}^{2}}[(\dfrac{r_{0}}{r})^{2n+2}-\dfrac{1}{4}(\dfrac{r_{0}}{r})^{3n+2}]\geq0
\end{equation}
From equation (\ref{rho}), we can conclude that the positive energy density condition is satisfied for  $ 0<n<1 $ (see Fig. \ref{fig10}). Furthermore, expressions (\ref{pr}) and (\ref{pt}) are negative in this range (see Fig. \ref{fig11}).
\begin{table}
\begin{center}
\setlength{\tabcolsep}{1cm}
\begin{tabular} {| c | c | c | c | c |}
\hline
Flare out&$\cellcolor{red}$ &$\cellcolor{red}$ &     &   \\ \hline
$\rho$ &&&&$\cellcolor{red}$  \\
\hline
$ \zeta >0$ & $\cellcolor{red}$ &$\cellcolor{red}$& $\cellcolor{red}$ &$$  \\
 \hline
n &-2 and less&-1&+1&+2 and more\\
\hline
\end{tabular}
\end{center}
\caption{The status of wormhole conditions versus the key parameter $ n $. Shaded areas correspond to the violation of each condition. \label{tab:tank}}
\end{table}

\section{Conclusion}
In this paper, we introduced a new wormhole solution of Einstein equations with a multi-polytropic source. With the help of wormhole conditions (flare out, asymptotic flatness, positive mass), the parameter of the model was constrained ($ 0<n<1 $). Full lensing (which includes both weak and strong lensing) were calculated. It was shown that the behavior of this object is similar to the lensing behavior of a black hole at least far enough from the throat. Also the light curves of microlensing of the positive mass (but violating energy condition) of the extended lens was shown to be similar to a point source object. In oredr have a finite mass, we used the thin shell formalism, and investigated the behavior of light ray deflection by using the Fermat principle and junction conditions. Then, we studied the raidal stability of the shell. Also, we learned that within the multi-polytropic $ EoS $, static wormholes can respect the positive energy density condition all over the space. By plotting $ \rho+P_{r} $  and $\rho+P_{t}  $ versus  $ r$, we found that these quantities are negative everywhere, thus violating the WEC.

\pagebreak

 \end{document}